\documentclass[10pt,conference,compsocconf]{IEEEtran}
\usepackage{times}
\usepackage{slashbox}

\usepackage{caption}
\ifCLASSINFOpdf
\usepackage[pdftex]{graphicx}
\else
\usepackage[dvips]{graphicx}
\fi
\begin{document}
\pagenumbering{gobble}
%
\title{\textbf{\Large Effective Self-Healing Networks
against Attacks or\\ Disasters in Resource Allocation Control}}

\author{\IEEEauthorblockN{~\\[-0.4ex]\large Yukio Hayashi\\[0.3ex]\normalsize}
\IEEEauthorblockA{Graduate School of Advanced\\
Institute of Science and Technology,\\
Japan Advanced Institute of\\
Science and Technology\\
Nomi-city, Ishikawa 923-1292\\
Email: {\tt yhayashi@jaist.ac.jp}}
\and
\IEEEauthorblockN{~\\[-0.4ex]\large Atsushi Tanaka\\[0.3ex]\normalsize}
\IEEEauthorblockA{Department of Informatics and Electronics\\
Faculty of Engineering,\\
Yamagata University\\
Yonezawa-city, Yamagata 992-8510\\
Email: {\tt tanaka@yamagata-u.ac.jp}}
\and
\IEEEauthorblockN{~\\[-0.4ex]\large Jun Matsukubo\\[0.3ex]\normalsize}
\IEEEauthorblockA{Department of Creative Engineering,\\
National Institute of Technology\\
Kitakyushu College\\
Kitakyushu-city, Fukuoka 802-0985\\
Email: {\tt jmatsu@apps.kct.ac.jp}}
}


%


\maketitle

\begin{abstract}
With increasing threats by large attacks or disasters, 
the time has come to reconstruct network infrastructures 
such as communication or transportation systems 
rather than to recover them as before in case of accidents, 
because many real networks are extremely vulnerable.
Thus, we consider self-healing mechanisms by rewirings 
(reuse or addition of links) 
to be sustainable and resilient networks 
even against malicious attacks.
In distributed local process for healing, 
the key strategies are 
the extension of candidates of linked nodes and 
enhancing loops by applying a message-passing algorithm 
inspired from statistical physics.
Simulation results show that 
our proposed combination of ring formation and enhancing loops 
is particularly effective in comparison with the conventional methods, 
when more than half damaged links alive or are compensated 
from reserved ones.
\end{abstract}



%
\IEEEpeerreviewmaketitle

\section{Introduction}

In contemporary world, 
network infrastructures, such as communication, trading, 
transportation, energy and water supply systems 
are crucial for supporting social activities, economy, 
industrial production, etc., 
while increasing the frequency of large disasters or military conflicts 
turn threats by destroying the functions into reality. 
To confront the serious problems, 
a new supple approach {\it resilience} \cite{Zolli12,Hollnagel06} 
attracts much attention in system engineering, biology, 
ecology, and sociology. 
Resilience means the ability to sustain basic objective and integrity 
even in encountering with the extreme change of situations or 
environments (e.g., by disasters or malicious attacks) 
for technological system, organization, or individual \cite{Zolli12}. 
However, to be resilient system, 
the concept of safety against accidents or disruptions 
should be extended from {\it Safety-I} to {\it Safety-II} \cite{Hollnagel14}:
from ``as few things as possible go wrong'' 
to ``as many things as possible go right'', 
from ``reactive, respond when something happens'' 
to ``proactive, continuously trying to anticipate developments and events'', 
and so on, in these two complementary views, which do not conflict. 
Safety-II requires to adjust, adapt, develop, and design better 
processes with technological or human resource allocations. 
Moreover, 
the concept of resilience includes reorganization or 
reconstruction of the system 
with adaptive capacity beyond the conventional recovery \cite{Folke06}, 
as shown in Table \ref{table_adaptove_capacity}.
Such paradigm shift 
has affinity to self-adaptive mechanism or system. 
Thus, 
we focus on 
developing new mechanism to lead to (social-ecological) resilience 
with reconstruction as far as accepting innovation rather 
than finding and decreasing weak or wrong parts in a network 
system. 

In this paper, we study how to reconstruct a sustainable network 
under limited resource, and propose effective self-healing methods 
based on enhancing loops through a local process around damaged parts.
The motivations for enhancing loops are as follows.
There is a common topological structure called Scale-Free (SF) 
in many social, biological, and technological networks 
\cite{Amaral00,Barabasi03}.
Although the SF networks have an extreme vulnerability against 
malicious attacks \cite{Albert00}, 
it has been found that onion-like structure 
with positive degree-degree correlations gives the optimal 
robustness of connectivity \cite{Schneider11,Tanizawa12}.
Onion-like structure can be generated by 
whole rewiring \cite{Schneider11,Holme11} 
in enhancing the correlations under a given degree distribution. 
Moreover, 
since dismantling and decycling problems are asymptotically 
equivalent 
at infinite graphs in a large class of random networks 
with light-tailed degree distribution \cite{Braunstein16}, 
a tree remains without loops 
at the critical state before 
the complete fragmentation by node removals.
Dismantling (or decycling) problem known as NP-hard \cite{Karp72} 
is to find the minimum set of nodes in which removal leaves 
a graph with the largest connected cluster 
whose size is at most a constant (or a graph without loops). 
It is suggested that 
the robustness becomes stronger as many loops exist as possible.
In fact, to be onion-like networks, 
enhancing loops by copying or intermediation 
is effective for improving the robustness 
in incrementally growing methods  
\cite{Hayashi14,Hayashi18a} based on a local distributed process. 
Thus, we remark that 
loops make bypasses and may be more important 
than the degree-degree correlations 
in order to improve the connectivity in a network reconstruction 
after large disasters or attacks.
However, identifying the necessary nodes to form loops is intractable 
due to combinatorial NP-hardness, 
we effectively apply an approximate calculation based on a 
statistical physics approach in our proposed self-healing.
We assume that rewirings (reuse of undestroyed links) 
are performed by changing directions or ranges 
of flight routes or wireless beams in the healing process, 
though we do not discuss the detail realization 
that depends on the current or future technologies and target systems.


\captionsetup{font={footnotesize,sc},justification=centering,labelsep=period}%
\begin{table}[htbp]
\caption{A sequence of resilience concepts which are partially extracted from 
\cite{Folke06}.}\label{table_adaptove_capacity}
\centering%
\begin{tabular}{lll}
\hline
\textit{Resilience concept} & \textit{Characteristics} & \textit{Focus on} \\
\hline
Engineering resilience & Return time, & Recovery,\\
                       & efficiency    & constancy\\ 
 & & \\
Ecological/ecosystem   & Buffer capacity, & Persistence,\\
resilience & withstand shock,  & robustness\\
           & maintain function & \\ 
 & & \\
Social$-$ecological & Interplay disturbance & {\bf Adaptive capacity},\\
resilience        & and {\bf reorganization},   & transformability,\\
                  & sustaining and        & learning,\\
                  & developing            & innovation\\
\hline
\end{tabular}
\end{table}
\captionsetup{font={footnotesize,rm},justification=centering,labelsep=period}%

\section{Related Works}
We briefly review recent progress of typical methods for recovery and 
healing of a network in complex network science (inspired from fractal 
statistical physics) and computer science.

In complex network science, 
several recovery and healing methods have been proposed. 
As one of the recovery methods, 
the strategies of random, 
greedy (for regaining the largest connectivity), and preferential 
recovery weighted by population have been considered 
in taking into account the order of recovered links \cite{Hu16}.
The effectiveness of recovery from localized attacks
is investigated on a square lattice. 
Against link failures, 
a simple recovery method has been also introduced 
to reconstruct an active tree for delivering from a source node 
by using back-up links \cite{Scala14}.
However, it is unclear 
which pairs of two nodes should be prepared for back-up links 
in advance. 

On the other hand, 
a self-healing method has been proposed by establishing new random 
links on interdependent (two-layered) networks of square lattices 
\cite{Kertesz14}, and the effect against node attacks 
is numerically studied. 
In particular, for adding links by the healing process, 
the candidates of linked nodes are incrementally extended from 
only the direct neighbors of the removed node by attacks 
until no more separation of components occurs. 
In other words, 
the whole connectivity is maintained except the isolating 
removed parts (known as induced sub-graphs for removed nodes 
in computer science). 
Note that such an extension of the candidates of linked nodes
is a key idea in our proposed self-healing method as mentioned later. 

Furthermore, 
the following self-healing methods, 
whose effects are investigated for some data of real networks, 
are worthy to note.
One is a distributed local repair in order of a priority to 
the most damaged node \cite{Gallos15}.
In the repair by linking 
from the most damaged node to a randomly chosen node from the 
unremoved node set in its next-nearest neighbors before attacks, 
the order of damaged nodes is according to 
the smaller fraction $k_{dam} / k_{orig}$ of its remained degree $k_{dam}$ 
and the original degree $k_{orig}$ before the attacks.
The selections are repeated 
until reaching a given rate $f_{s}$ controlled by 
the fraction of nodes whose $k_{dam} / k_{orig}$ falls below a threshold. 
Another is a bypass rewiring \cite{Park16} 
on more limited resource of links (wire cables, wireless communication 
or transportation lines between two nodes) and 
ports (channels or plug sockets at a node). 
To establish links between pair nodes, 
a node is randomly chosen only one time in the neighbors of 
each removed node. 
When $k_{i}$ denotes the degree of removed node $i$, 
only $\lfloor k_{i}/2 \rfloor$ links are reused.
Note that a degree represents the number of using 
ports at the node.
In the bypass rewiring, 
reserved additional ports are not necessary: 
they do not exceed the original one before attacks. 
Moreover, greedy bypass rewiring \cite{Park16} 
is proposed in order to improve the 
robustness, the selection of pair nodes is based on the number 
of the links not yet rewired and the size of the neighboring components.

In computer science, 
ForgivingTree algorithm has been proposed \cite{Hayes08}.
Under the repeated attacks, 
the following self-healing is processed one-by-one after each 
node removal, except when the removed node is a leaf 
(whose degree is one).
It is based on both distributed process of sending messages 
and data structure, furthermore developed to an efficient 
algorithm called as compact routing \cite{Castaneda18}. 
In each rewiring process, 
a removed node and its links are replaced by a binary tree. 
Note that each vertex of the binary tree was the neighbors 
of the removed node, whose links to the neighbors are reused 
as the edges of the binary tree. 
Thus, additional links for healing is unnecessary. 
It is remarkable for computation 
(e.g. in routing or information spreading) that 
the multiplicative factor of diameter of the graph after healing 
is never more than 
$O( \log k_{max})$, where $k_{max}$ is the maximum degree in 
the original network, because of the replacing by binary 
trees.
However, 
the robustness of connectivity is not taken into account 
in the limited rewiring based on binary trees, 
since a tree structure is easily disconnected into subtrees 
by any attack to the joint node.
In other research, 
a recovery strategy with resource allocation of bandwidth 
in a communication network is discussed 
at several service levels (from full to partial service) 
w.r.t what and how optimization \cite{Savas14}, 
although considering the link's thickness (e.g. defined by 
bandwidth or transportation amounts) is out of our current 
scope. 

The characteristics of resource allocation are summarized in Table 
\ref{table_resource} for the above conventional 
and our proposed methods, 
although there has been no discussion about resource of links and ports.
ForgivingTree or bypass rewiring methods is not controllable but 
strongly  depending on the reuse of all or half links before attacks.
We assume that 
some links emanated from a removed node $i$ 
can be reused for healing 
by local rewiring between the neighbors.
Some links (cable lines) may work at the neighbor's sides, 
even though they are disconnected at the removed node's side.
As a control parameter in our simulation, we set 
the reusable rate $r_{h}$ according to the damage, 
on the assumption $k_{i} (1 - r_{h})$ links do not work in 
the removed node's degree $k_{i}$. 
In the two kinds of resource, 
we consider that ports work independently from connection links, 
as similar to a relation of airport runaway (or plug socket) 
and flight by airplane (or cable line).

\captionsetup{font={footnotesize,sc},justification=centering,labelsep=period}%
\begin{table}[htbp]
\caption{Reserved resource at a node in self-healing methods}\label{table_resource}
\centering%
\begin{tabular}{lll}
\hline
\textit{Method} & \textit{Additional links} & \textit{Additional ports} \\
\hline
ForgivingTree \cite{Hayes08} & Unnecessary, & Two or three at most \\
             & enough by the original & in a binary tree \\
             & under the assumption of reuse & \\ \hline
Bypass Rewiring \cite{Park16} &  Unnecessary, & Unnecessary, \\
             & if about half is reusable & enough by the original \\ 
             & from the original & \\ \hline
Simple Local Repair \cite{Gallos15} & Controllable & Necessary \\
             & $f_{s}(1-q)N$ & according to $f_{s}$ \\
             & & and attack rate $q$ \\ \hline
Our Proposed Method & Controllable & Necessary \\
             & $M_{h}$  & according to $r_{h}$ \\
             & & and attack rate $q$ \\ 
\hline
\end{tabular}
\end{table}
\captionsetup{font={footnotesize,rm},justification=centering,labelsep=period}%

\section{Effective Self-Healing}
\subsection{Outline of Proposed Methods}
We assume that almost simultaneously attacked nodes 
are not recoverable immediately, 
therefore are removed from the network function for a while.
In case of emergency for healing, 
unconnected two nodes are chosen and rewired as the 
reconstruction assistance or reuse of links 
emanated from removed $q N$ nodes, 
when the fraction of attacks is $q$. 
The healing process in each of the following 1), 2), and 3) 
is initiated just after detecting attacks and repeated by 
$M_{h} \stackrel{\rm def}{=} r_{h} \times \tilde{\sum}_{i \in D_{q}} k_{i}$ links. 
Here, 
$\tilde{\sum}_{i \in D_{q}} k_{i}$ means the number 
without multiple counts of lost links by attacks. 
$D_{q}$ denotes the set of removed nodes, $| D_{q} | = q N$.
The key strategies are 
1) enhancing loops contributes to improve the robustness 
\cite{Hayashi18a,Hayashi18b}, 
2) forming rings that encloses damaged parts 
is able to maintain the 
connectivity on the edges of extended neighbors, and 
3) complementary effects of 1) and 2) in the limited resource 
of $M_{h}$ links. 
Any one of them is performed as the healing process.

\begin{figure}[htb]
  \centering
  \includegraphics[width=0.8\linewidth]{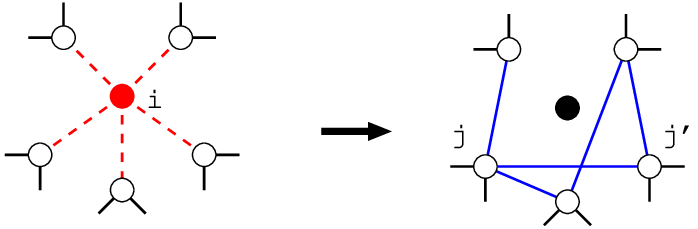}
\caption{Rewiring between nodes with small $q_{j}^{0} + q_{j'}^{0}$.}
\label{fig_smaller_q}
\end{figure}

\begin{figure}[htb]
  \centering
  \includegraphics[width=1.0\linewidth]{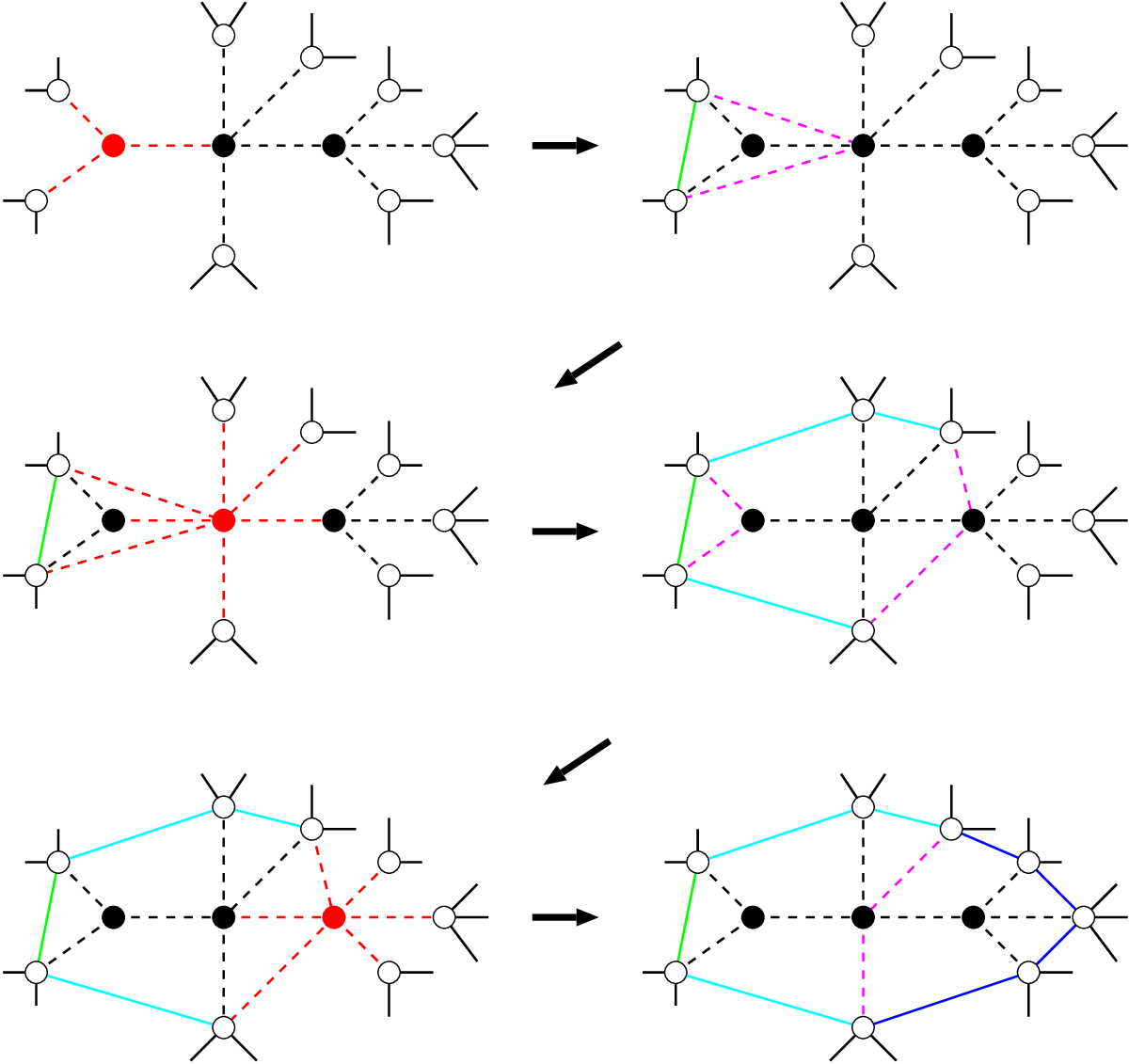}
\caption{Generation process of a ring.}
\label{fig_repair_ring}
\end{figure}

\begin{description}
  \item[1)] Enhancing loops for smaller $q_{j}^{0} + q_{j'}^{0}$ \\
    To select two nodes in the neighbor nodes $j, j' \in \partial i$ 
    in the increasing order of $q_{j}^{0} + q_{j'}^{0}$ for all $i \in D_{q}$, 
    as shown in Fig. \ref{fig_smaller_q}.
  \item[2)] Extended ring\\ To make rings of simple cycles without crossing, 
    the neighbors are extended from the first damaged, the second 
    damaged, $\ldots$, to the last damaged area in this order, 
    as shown in Fig. \ref{fig_repair_ring}.
  \item[3)] Combination of extended rings \& enhancing loops for smaller $q_{j}^{0} + q_{j'}^{0}$\\
    After using $M_{r} \leq \tilde{\sum}_{i \in D_{q}} k_{i}$ links for the rings,
    if $M_{h} > M_{r}$ \footnote{If $M_{h} < M_{r}$, 
      a ring is incomplete and an open chain is generated among 
      the extended neighbors. In this case, additional rewirings 
      based on smaller $q_{j}^{0} + q_{j'}^{0}$ are not performed 
      due to lack of links.}
    then the selections of two nodes in the extended neighbors on ring 
    are repeated in the increasing order of $q_{j}^{0} + q_{j'}^{0}$ 
    for $M_{h}- M_{r}$ links. 
\end{description}

Enhancing loops is performed by applying the values of $q_{i}^{0}$ 
(introduced in next subsection) for estimating Feedback Vertex Set 
(FVS) whose nodes are necessary to form loops.
Since a node $i$ with small $q_{i}^{0}$ belongs to a dangling subtree 
with high probability, by connecting such nodes, it is expected 
that a new loop on which a part of the subtree is included is added. 
From left to right in Fig. \ref{fig_smaller_q}, 
the original red links emanated from the removed node $i$ (marked by 
filled circle) are reused as the blue ones for the healing.
When there is at least a path between the nodes $j$ and $j'$ in 
Fig. \ref{fig_smaller_q}, a new loop is created.
Note that the attacked node $i$ is isolatedly 
removed as breakdown. 

A ring is generated as follows.
In Fig. \ref{fig_repair_ring}, the process is initiated 
in order of removals of three nodes from left to right. 
Filled and open circles denote removed and active nodes, 
red and magenta lines denote removed and virtually added links,
respectively. 
From top left to top right in Fig. \ref{fig_repair_ring}, 
a red node and its links are damaged, a ring formation around 
the 1st removal node at the left 
is tried to the direct neighbors of it. 
A green link is established, 
while virtual magenta links are considered by sending messages 
to active neighbors. 
From middle left to middle right in Fig. \ref{fig_repair_ring}, 
the ring formation around the 2nd removal node at the center 
is tried again to the neighbors which include 
the extended ones by the virtual links. 
Light blue links are added, 
but virtual magenta links are considered. 
From bottom left to bottom right in Fig. \ref{fig_repair_ring}, 
the ring formation around the 3rd removal node at the right 
is tried similarly. 
Finally, a ring is established by green, light blue, and blue links. 
The connections between neighbors on a ring are in random order 
except through the extension process.

\subsection{Applying Belief Propagation Algorithm}
We review the following approximation algorithm \cite{Zhou13,Zhou16} 
derived for estimating FVS known as NP-hard problem \cite{Karp72}.
It is based on a cavity method in statistical physics 
in the assumption that nodes $j \in \partial i$ 
are mutually independent of each other when node $i$ is removed. 
The joint probability is 
${\cal P}_{\backslash i}({A_{j}: j \in \partial i}) 
  \approx \Pi_{j \in \partial i} q^{A_{j}}_{j \rightarrow i}$ 
by the product of independent marginal probability 
$q^{A_{j}}_{j \rightarrow i}$ 
for the state $A_{j}$ as the index of $i$'s root. 
In the cavity graph, 
if all nodes $j \in \partial i$ are either empty ($A_{j} = 0$)
or roots ($A_{j} = j$), 
the added node $i$ can be a root ($A_{i} = i$).
There are the following exclusive states.
\begin{enumerate}
  \item $A_{i} = 0$: $i$ is empty (removed). 
    Since $i$ is unnecessary as a root, it belongs to FVS.
  \item $A_{i} = i$: $i$ becomes its own root. 
    The state $A_{j} = j$ of $j \in \partial i$ 
    is changeable to $A_{j} = i$ when node $i$ is added.
  \item $A_{i} = k$: one node $k \in \partial i$ becomes the root
    of $i$ when it is added, 
    if $k$ is occupied and all other $j \in \partial i$ 
    are either empty or roots.
\end{enumerate}
The corresponding probabilities to the above three states are represented by 
\begin{equation}
  q^{0}_{i} \stackrel{\rm def}{=} \frac{1}{z_{i}(t)}, \label{eq_BP1}
\end{equation}
\[
  q^{i}_{i} \stackrel{\rm def}{=} 
  \frac{e^{x} \Pi_{j \in\partial i(t)} 
  \left[ q^{0}_{j \rightarrow i} + q^{j}_{j \rightarrow i} \right]}{z_{i}(t)},
\]
\[
 q^{k}_{i} \stackrel{\rm def}{=} 
 \frac{e^{x} \frac{(1 - q^{0}_{k \rightarrow i})}{
        q^{0}_{k \rightarrow i} + q^{k}_{k \rightarrow i}} 
  \Pi_{j \in\partial i(t)} 
  \left[ q^{0}_{j \rightarrow i} + q^{j}_{j \rightarrow i} \right]}{z_{i}(t)},
\]
\begin{equation}
  q^{0}_{i \rightarrow j} = \frac{1}{z_{i \rightarrow j}(t)}, 
\end{equation}
\begin{equation}
  q^{i}_{i \rightarrow j} = 
  \frac{e^{x} \Pi_{k \in \partial i(t)\backslash j} 
    \left[ q^{0}_{k \rightarrow i} + q^{k}_{k \rightarrow i} \right]
  }{z_{i \rightarrow j}(t)}, 
\end{equation}
where $\partial i(t)$ denotes node $i$'s set of connecting neighbor nodes 
at time $t$, 
and $x > 0$ is a parameter of inverse temperature. 
The normalization constants are 
\begin{equation}
  z_{i}(t) \stackrel{\rm def}{=} 
  1 + e^{x} \left[ 1 + \sum_{k \in \partial i(t)} 
      \frac{1 - q^{0}_{k \rightarrow i}}{
        q^{0}_{k \rightarrow i} + q^{k}_{k \rightarrow i}} \right]
  \Pi_{j \in\partial i(t)} 
  \left[ q^{0}_{j \rightarrow i} + q^{j}_{j \rightarrow i} \right],
  \label{eq_BP4}
\end{equation}
\begin{eqnarray}
  z_{i \rightarrow j}(t) & \stackrel{\rm def}{=} &
  1 + e^{x} \Pi_{k \in \partial i(t)\backslash j} 
  \left[ q^{0}_{k \rightarrow i} + q^{k}_{k \rightarrow i} \right] \\
  & & \times \left[ 1 + \sum_{l \in \partial i(t)\backslash j} 
    \frac{1 - q^{0}_{l \rightarrow i}}{
      q^{0}_{l \rightarrow i} + q^{l}_{l \rightarrow i}} \right], 
  \label{eq_BP5}
\end{eqnarray}
to be satisfied for any node $i$ and link $i \rightarrow j$ as 
\[
  q^{0}_{i} + q^{i}_{i} + \sum_{k \in \partial i} q^{k}_{i} = 1,
\]
\[
  q^{0}_{i \rightarrow j} + q^{i}_{i \rightarrow j} 
  + \sum_{k \in \partial i} q^{k}_{i \rightarrow j} = 1.
\]
The message-passing iterated by Eqs. (\ref{eq_BP1})-(\ref{eq_BP5}) 
is called belief propagation (BP). 
These calculations of $q_{i}^{0}$, $q_{i}^{i}$, $q_{i}^{k}$, 
$q_{i \rightarrow j}^{0}$, $q_{i \rightarrow j}^{i}$, and 
$q_{i \rightarrow j}^{k}$ are locally 
executed through the message-passing 
until to be self-consistent in principle but practically to 
reach appropriate rounds 
from initial setting of $(0,1)$ random values.
The unit time from $t$ to $t+1$ for calculating a set 
$\{ q^{0}_{i} \}$ consists of a number of rounds by 
updating Eqs. (\ref{eq_BP1})-(\ref{eq_BP5}) 
in order of random permutation of the total $N$ nodes.
The distributed calculations can be also considered.

\captionsetup{font={footnotesize,sc},justification=centering,labelsep=period}%
\begin{table}[htbp]
\caption{number of additional ports
in our proposed combination method.}\label{table_ports_proposed}
\centering
\begin{tiny}

Open Flight: $k_{max} = 242$
\begin{tabular}{c|ccccccccc}\hline
\backslashbox{\textit{$r_h$}}{\textit{$q$}}
 & \textit{0.1} & \textit{0.2} & \textit{0.3} & \textit{0.4} & \textit{0.5} & \textit{0.6} & \textit{0.7} & \textit{0.8} & \textit{0.9} \\ \hline
0.05 &  1.0 &  1.0 &  1.0 &  1.0 &  1.0 &  1.0 &  1.0 & 32.8 & 73.2 \\
 & ( 0.3) & ( 0.4) & ( 0.5) & ( 0.5) & ( 0.5) & ( 0.5) & ( 0.5) & ( 1.4) & ( 4.6) \\
0.1 &  1.0 &  1.0 &  1.0 &  1.0 & 18.2 & 80.8 & 194.1 & 246.7 & 203.5 \\
 & ( 0.4) & ( 0.5) & ( 0.5) & ( 0.5) & ( 0.7) & ( 1.4) & ( 2.3) & ( 4.1) & ( 9.0) \\
0.2 & 89.2 & 84.7 & 294.3 & 347.0 & 362.0 & 446.5 & 412.5 & 341.1 & 228.2 \\
 & ( 1.0) & ( 1.5) & ( 1.7) & ( 2.3) & ( 3.1) & ( 4.0) & ( 5.8) & ( 9.0) & (19.1) \\
0.5 & 253.1 & 244.0 & 636.4 & 671.6 & 726.6 & 584.1 & 539.8 & 407.9 & 233.0 \\
 & ( 4.8) & ( 6.3) & ( 6.3) & ( 7.5) & ( 9.2) & (11.8) & (15.9) & (24.3) & (50.8) \\
1.0 & 403.0 & 449.4 & 807.8 & 822.1 & 755.8 & 626.7 & 571.2 & 422.4 & 249.2 \\
 & (10.9) & (13.8) & (14.1) & (16.1) & (19.5) & (24.6) & (33.3) & (50.9) & (103.4) \\
\hline
\end{tabular}

\vspace{2mm}
As Oregon: $k_{max} = 1458$
\begin{tabular}{c|ccccccccc}\hline
\backslashbox{\textit{$r_h$}}{\textit{$q$}}
 & \textit{0.1} & \textit{0.2} & \textit{0.3} & \textit{0.4} & \textit{0.5} & \textit{0.6} & \textit{0.7} & \textit{0.8} & \textit{0.9} \\ \hline
0.05 &  1.0 &  1.0 &  1.0 &  1.0 &  1.0 &  1.0 &  1.0 &  1.0 &  1.0 \\
 & ( 0.4) & ( 0.5) & ( 0.5) & ( 0.5) & ( 0.5) & ( 0.6) & ( 0.6) & ( 0.5) & ( 0.5) \\
0.1 &  1.0 &  1.0 &  1.0 &  1.0 &  1.0 &  1.0 &  1.0 &  1.0 & 147.4 \\
 & ( 0.5) & ( 0.6) & ( 0.5) & ( 0.5) & ( 0.5) & ( 0.5) & ( 0.5) & ( 0.5) & ( 2.5) \\
0.2 &  1.0 &  1.0 &  1.0 &  1.0 &  1.0 &  1.0 & 176.9 & 440.1 & 473.7 \\
 & ( 0.5) & ( 0.5) & ( 0.5) & ( 0.5) & ( 0.5) & ( 0.5) & ( 1.1) & ( 2.4) & ( 6.1) \\
0.5 & 70.1 & 289.1 & 509.4 & 824.6 & 1112.4 & 1252.7 & 1122.7 & 820.1 & 485.1 \\
 & ( 0.7) & ( 1.0) & ( 1.3) & ( 1.8) & ( 2.4) & ( 3.3) & ( 4.9) & ( 8.1) & (17.7) \\
1.0 & 248.9 & 2275.3 & 2002.8 & 1801.2 & 1575.7 & 1343.0 & 1068.0 & 822.8 & 494.6 \\
 & ( 3.0) & ( 3.3) & ( 4.0) & ( 4.9) & ( 6.2) & ( 8.2) & (11.4) & (17.8) & (37.1) \\
\hline
\end{tabular}
\end{tiny}
\end{table}
\captionsetup{font={footnotesize,rm},justification=centering,labelsep=period}%

\captionsetup{font={footnotesize,sc},justification=centering,labelsep=period}%
\begin{table}[htbp]
\caption{number of additional ports
in the conventional simple local repair method.}\label{table_ports_pre92}
\centering
\begin{tiny}

Open Flight: $k_{max} = 242$
\begin{tabular}{c|ccccccccc}
\hline
\backslashbox{\textit{$r_h$}}{\textit{$q$}}
 & \textit{0.1} & \textit{0.2} & \textit{0.3} & \textit{0.4} & \textit{0.5} & \textit{0.6} & \textit{0.7} & \textit{0.8} & \textit{0.9} \\ \hline
0.05 & 14.9 & 8.1 & 7.7  & 3.7  & 3.7  & 4.1 & 3.8 & 3.2 & 2.3 \\
     & (3.2) & (1.7) & (1.3) & (1.2) & (1.3) & (1.3) & (1.3) & (1.3) & (1.2)\\
0.1  & 14.6 & 6.8 & 11.4 & 4.1  & 4.4  & 4   & 3.5  & 3.2 & 2.7 \\
     & (2.8) & (1.7) & (1.6) & (1.3) & (1.3) & (1.3) & (1.3) & (1.3) & (1.3)\\
0.2  & 4.8   & 4.6 & 4.4 & 4.1 & 4.2  & 3.9  & 3.7  & 3   & 2.7 \\
     & (1.3) & (1.4) & (1.3) & (1.3) & (1.3) & (1.3) & (1.3) & (1.2) & (1.2) \\
0.5  & 4.8  & 4.6 & 4.4  & 4.1  & 4.2  & 3.9 & 3.7  & 3   & 2.7 \\
     & (1.3) & (1.4) & (1.3) & (1.3) & (1.3) & (1.3) & (1.3) & (1.2) & (1.2) \\
1.0  & 4.8  & 4.6 & 4.4  & 4.1  & 4.2  & 3.9 & 3.7  & 3   & 2.7 \\
     & (1.3) & (1.4) & (1.3) & (1.3) & (1.3) & (1.3) & (1.3) & (1.2) & (1.2) \\ 
\hline
\end{tabular}

\vspace{2mm}
AS Oregon: $k_{max} = 1458$
\begin{tabular}{c|ccccccccc}
\hline
\backslashbox{\textit{$r_h$}}{\textit{$q$}}
 & \textit{0.1} & \textit{0.2} & \textit{0.3} & \textit{0.4} & \textit{0.5} & \textit{0.6} & \textit{0.7} & \textit{0.8} & \textit{0.9} \\ \hline
0.05 & 4.3 & 3.3 & 3.9 & 4.4 & 4   & 4.1 & 4.3 & 4.6 & 4.2 \\
     & (1.2) & (1.2) & (1.2) & (1.2) & (1.3) & (1.3) & (1.3) & (1.4) & (1.3)\\
0.1  & 5.3 & 4.6 & 4.8 & 4.6 & 5   & 5.1 & 4.7 & 4.6 & 4.8 \\
     & (1.3) & (1.3) & (1.3) & (1.3) & (1.4) & (1.4) & (1.4) & (1.4) & (1.4)\\
0.2  & 7   & 5.4 & 5.3 & 4.8 & 5.5 & 5   & 5   & 4.8 & 4.3 \\
     & (1.4) & (1.4) & (1.4) & (1.4) & (1.4) & (1.4) & (1.4) & (1.4) & (1.4)\\
0.5  & 5.4 & 5.9 & 5.8 & 5.7 & 5.5 & 5.2 & 5.4 & 5.1 & 4.3 \\
     & (1.4) & (1.4) & (1.4) & (1.4) & (1.4) & (1.4) & (1.4) & (1.4) & (1.4)\\
1.0  & 5.4 & 5.9 & 5.8 & 5.7 & 5.5 & 5.2 & 5.4 & 5.1 & 4.3 \\
     & (1.4) & (1.4) & (1.4) & (1.4) & (1.4) & (1.4) & (1.4) & (1.4) & (1.4)\\ \hline
\end{tabular}
\end{tiny}
\end{table}
\captionsetup{font={footnotesize,rm},justification=centering,labelsep=period}%

\section{Simulation Results}
We evaluate the effect of healing by two measures: 
the ratio $\frac{S(q)}{(1-q)N}$ \cite{Gallos15} for the 
connectivity and the efficiency 
$E \stackrel{\rm def}{=} \frac{1}{N(N-1)} 
\sum_{i \neq j} \frac{1}{L_{ij}}$, 
where $S(q)$ and $L_{ij}$ denote 
the size of GC (giant component or largest connected cluster) 
and the length of the shortest path counted by hops between 
$i$-$j$ nodes, respectively, 
for a network after removing $q N$ nodes by attacks 
with recalculation of the highest degree node as the target.
We investigate them 
for Open Flight between airports and Internet AS Oregon 
as examples of real networks \cite{net_data}, 
whose number of nodes and links are 
$N = 2905$, $M = 15645$, and $N = 6474$, $M = 12572$.
The following results are averaged over $10$ samples with 
random process for tie-breaking in a node selection or 
ordering of nodes on a ring.

\begin{figure}[htb]
\begin{minipage}{.49\textwidth}
\centering
  \includegraphics[width=.67\textwidth,angle=-90]{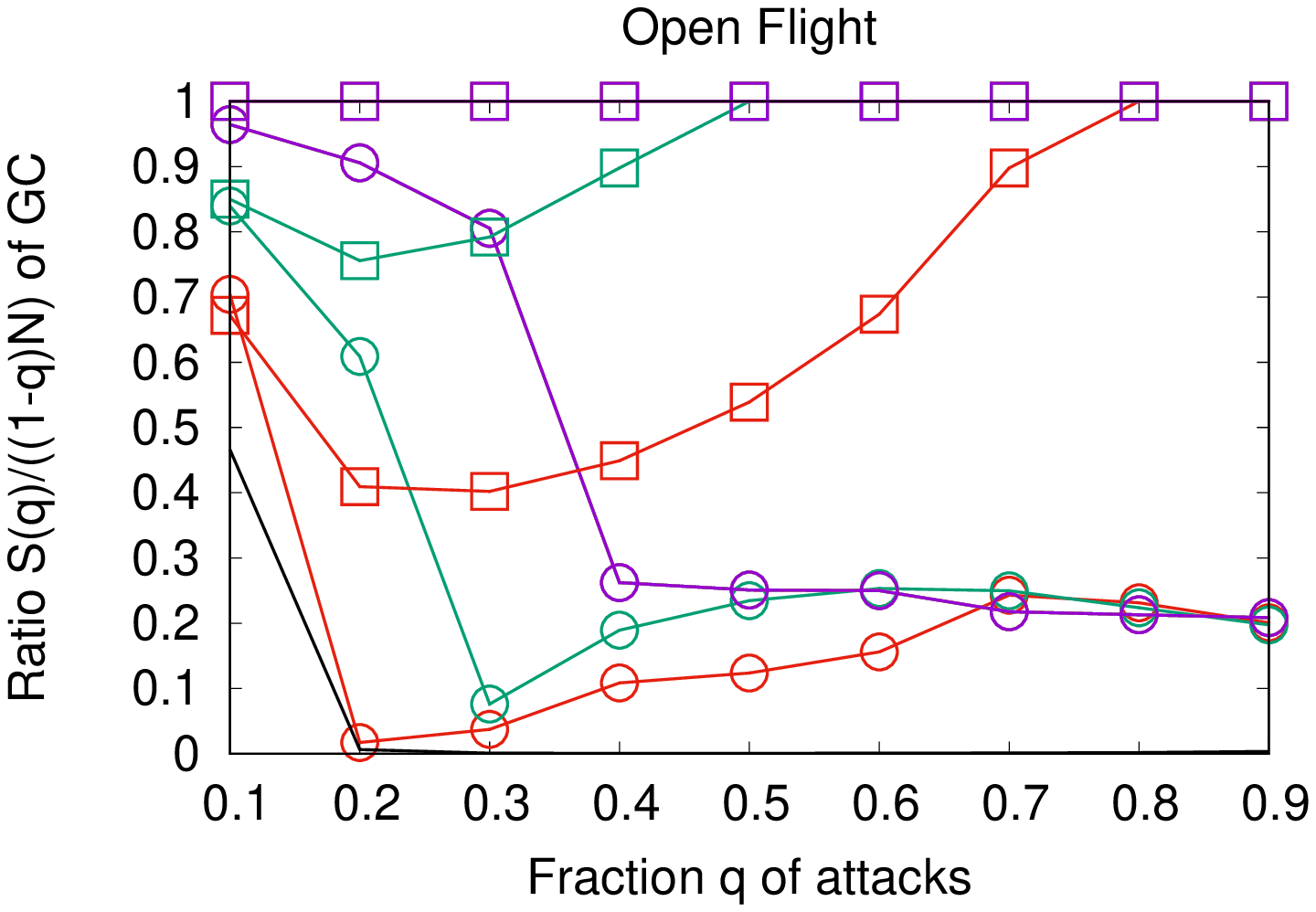}
\end{minipage}
\begin{minipage}{.49\textwidth}
\centering
  \includegraphics[width=.67\textwidth,angle=-90]{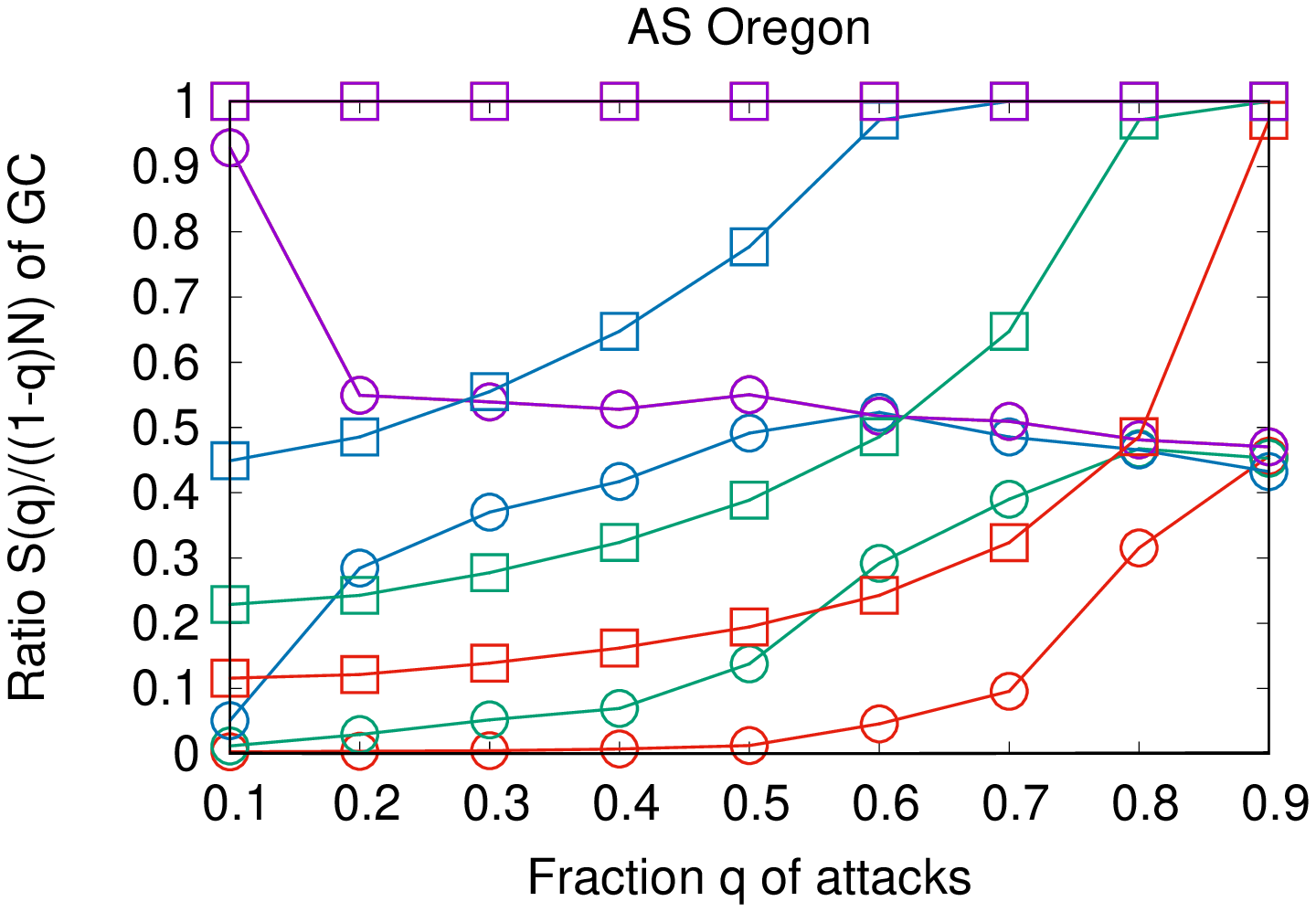}
\end{minipage}
\caption{Communicable or transportable size with healing by 
our proposed combination (square) 
and conventional simple local repair (circle).}
\label{fig_compare_GC}
\end{figure}

\begin{figure}[htb]
\begin{minipage}{.49\textwidth}
\centering
  \includegraphics[width=.67\textwidth,angle=-90]{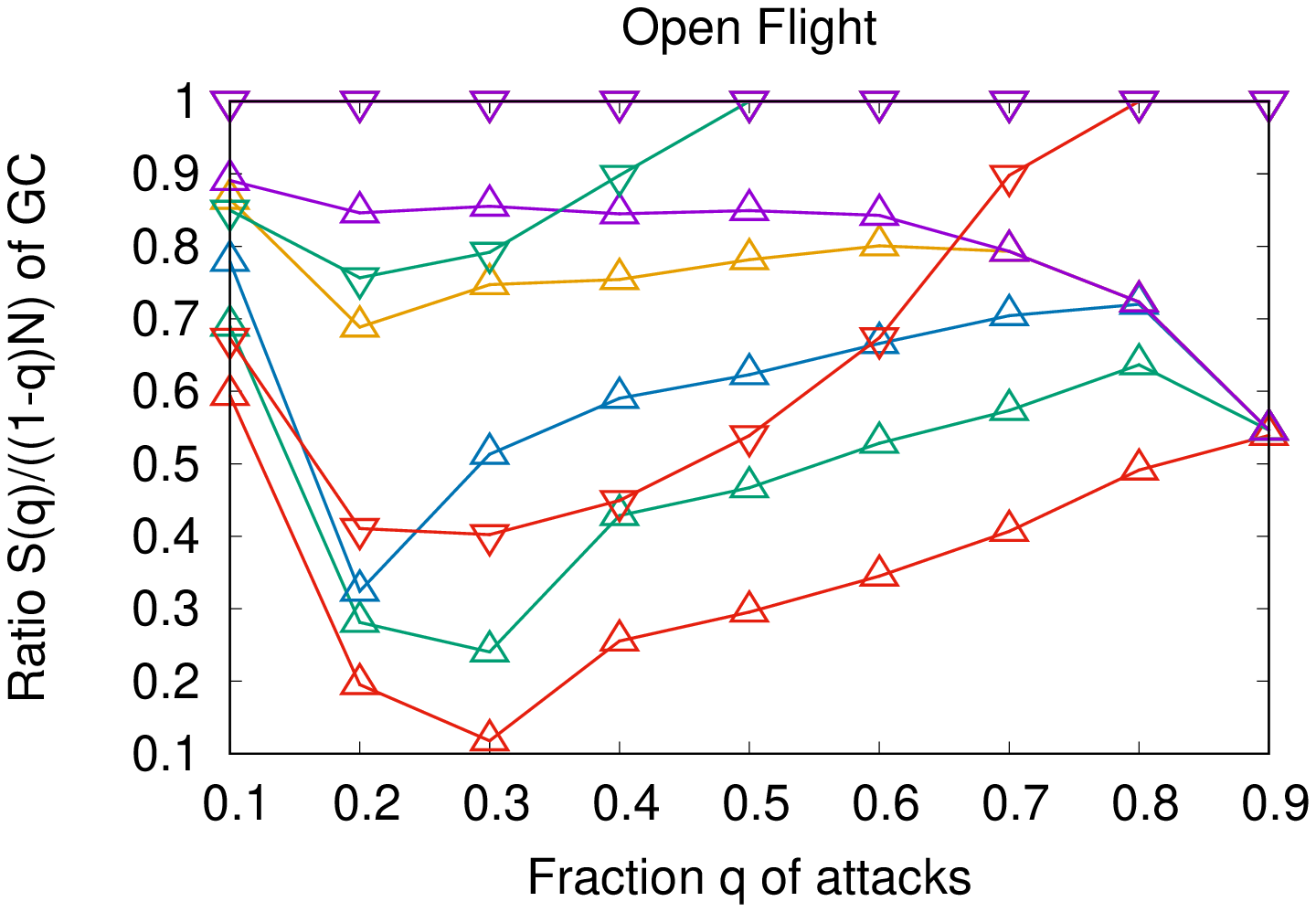}
\end{minipage}
\begin{minipage}{.49\textwidth}
\centering
  \includegraphics[width=.67\textwidth,angle=-90]{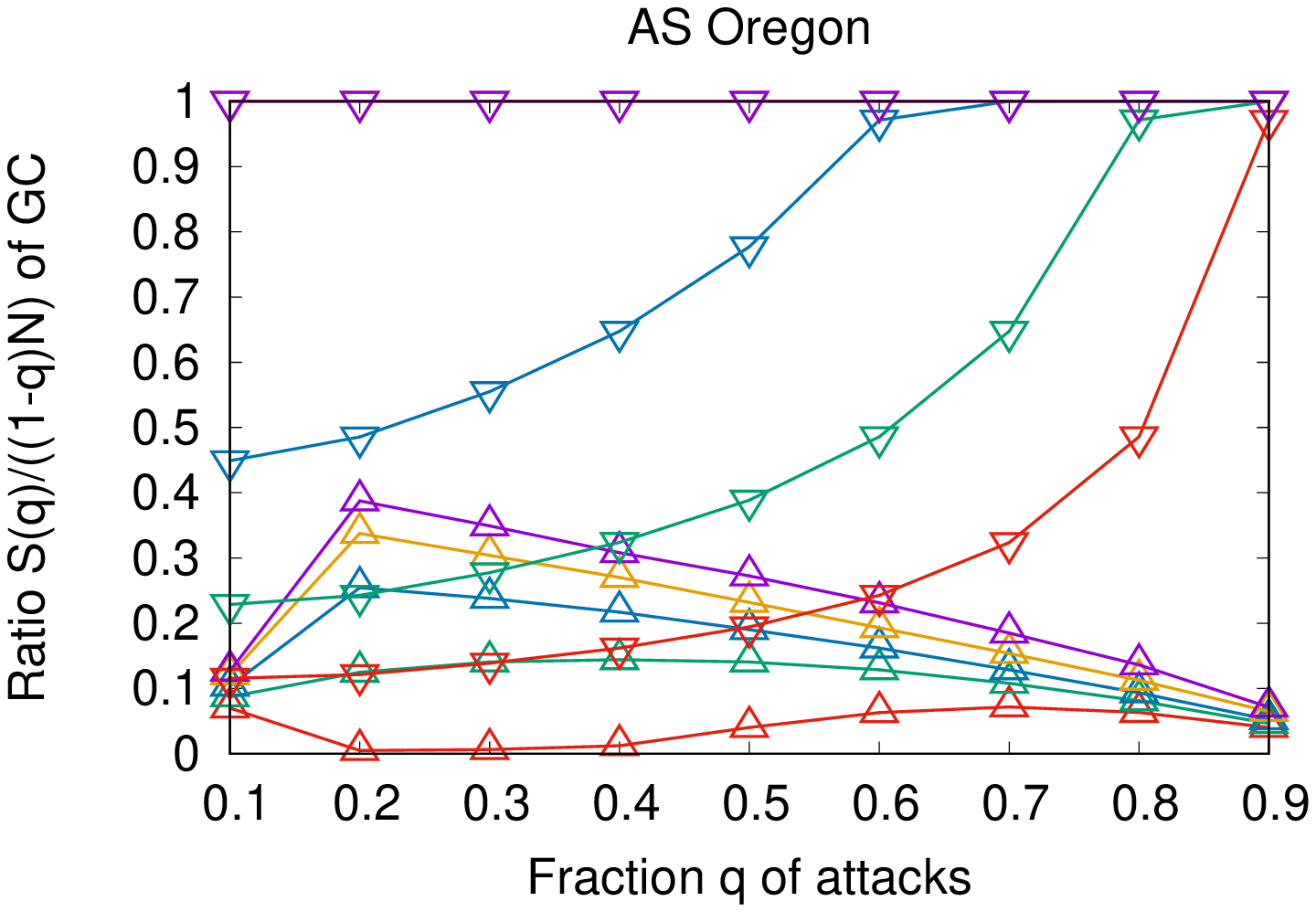}
\end{minipage}
\caption{Communicable or transportable size with healing by 
only enhancing loops (up triangle) 
or extended ring (down triangle).}
\label{fig_only_GC}
\end{figure}

Figure \ref{fig_compare_GC} shows the ratio of GC 
in the surviving nodes which may be divided into 
isolated clusters after attacks.
The number $M_{h}$ of rewiring is controlled 
by a parameter $r_{h}$ in the healing.
Red, green, blue, orange, and purple lines correspond to 
$r_{h} = 0.05$, $0.1$, $0.2$, $0.5$, and $1.0$.
Black line shows the result for no healing.
In comparison with same color lines, 
our proposed combination method (marked by square)
of extended ring and enhancing loops 
is superior with higher ratio than the conventional 
simple local repair \cite{Gallos15} method (marked by circle), 
whose healing works only for weak attacks in small $q$.
We remark that in our proposed combination method 
the cases of $r_{h} \geq 0.5$ (overlapped orange and purple lines 
marked by square) maintains the almost whole 
connectivity in the surviving $(1 - q)N$ nodes.
In other words, 
the network function can be revived completely, 
if more than half of links emanated from removed nodes 
are active. 
In $r_{h} \leq 0.1$ (green and red lines marked by square), 
making a ring is unfinished, the ratio is dropped.
Moreover, 
since the ratio in Fig. \ref{fig_only_GC} is lower than the ratio 
marked by square in Fig. \ref{fig_compare_GC}, 
only enhancing loops (marked by up-pointing triangle) or 
extended ring (marked by down-pointing triangle) has weaker effect 
than the combination. 
However, 
enhancing loops increase the ratio of GC moderately 
in $r_{h} \leq 0.1$ for $q \leq 0.5$
(green and red lines marked by up-pointing triangle) 
in Fig. \ref{fig_only_GC}.

\begin{figure}[htb]
\begin{minipage}{.49\textwidth}
\centering
  \includegraphics[width=.67\textwidth,angle=-90]{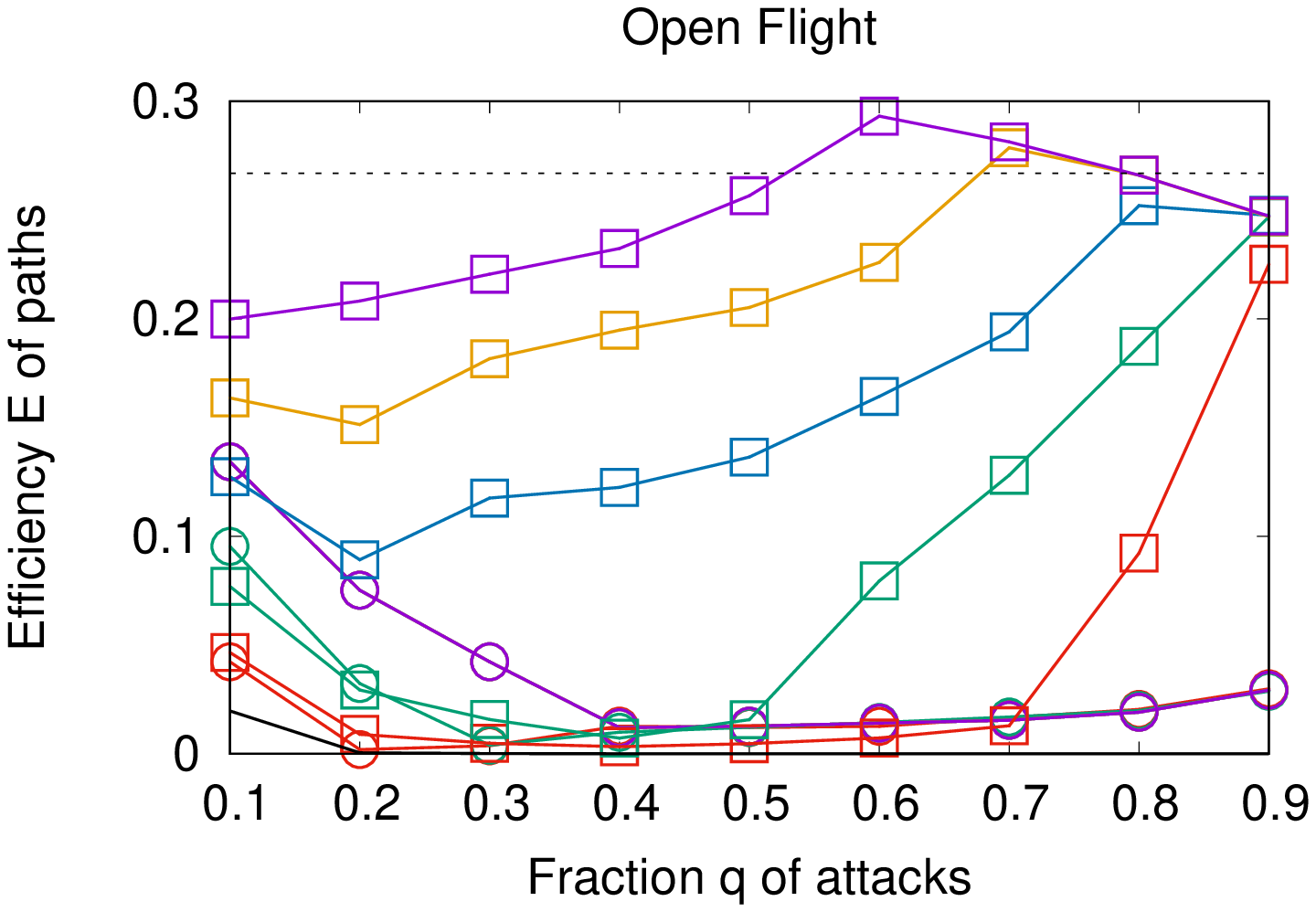}
\end{minipage}
\begin{minipage}{.49\textwidth}
\centering
  \includegraphics[width=.67\textwidth,angle=-90]{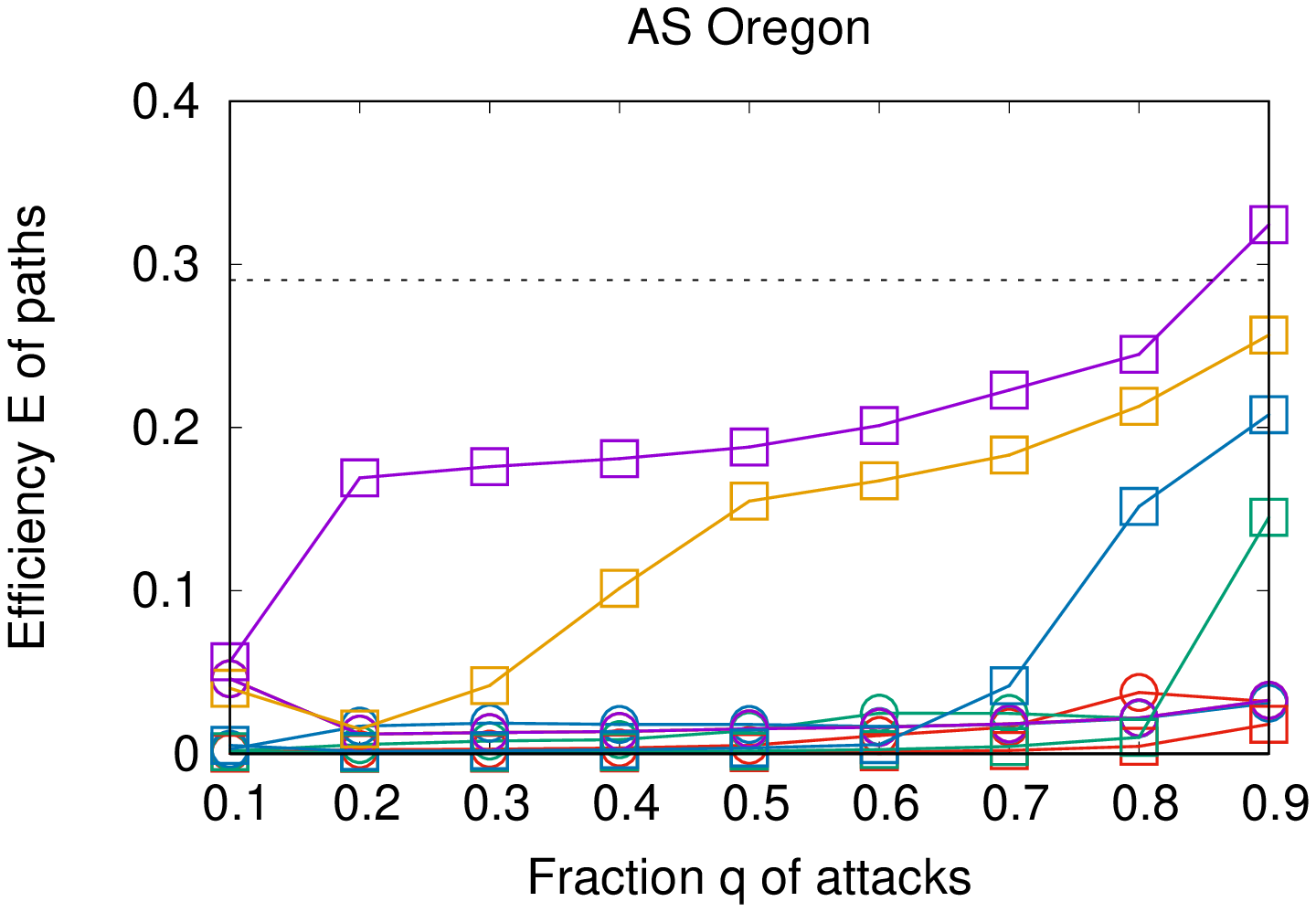}
\end{minipage}
\caption{Efficiency in the network with healing by 
our proposed combination (square) 
and conventional simple local repair (circle).}
\label{fig_compare_E}
\end{figure}

\begin{figure}[htb]
\begin{minipage}{.49\textwidth}
\centering
  \includegraphics[width=.67\textwidth,angle=-90]{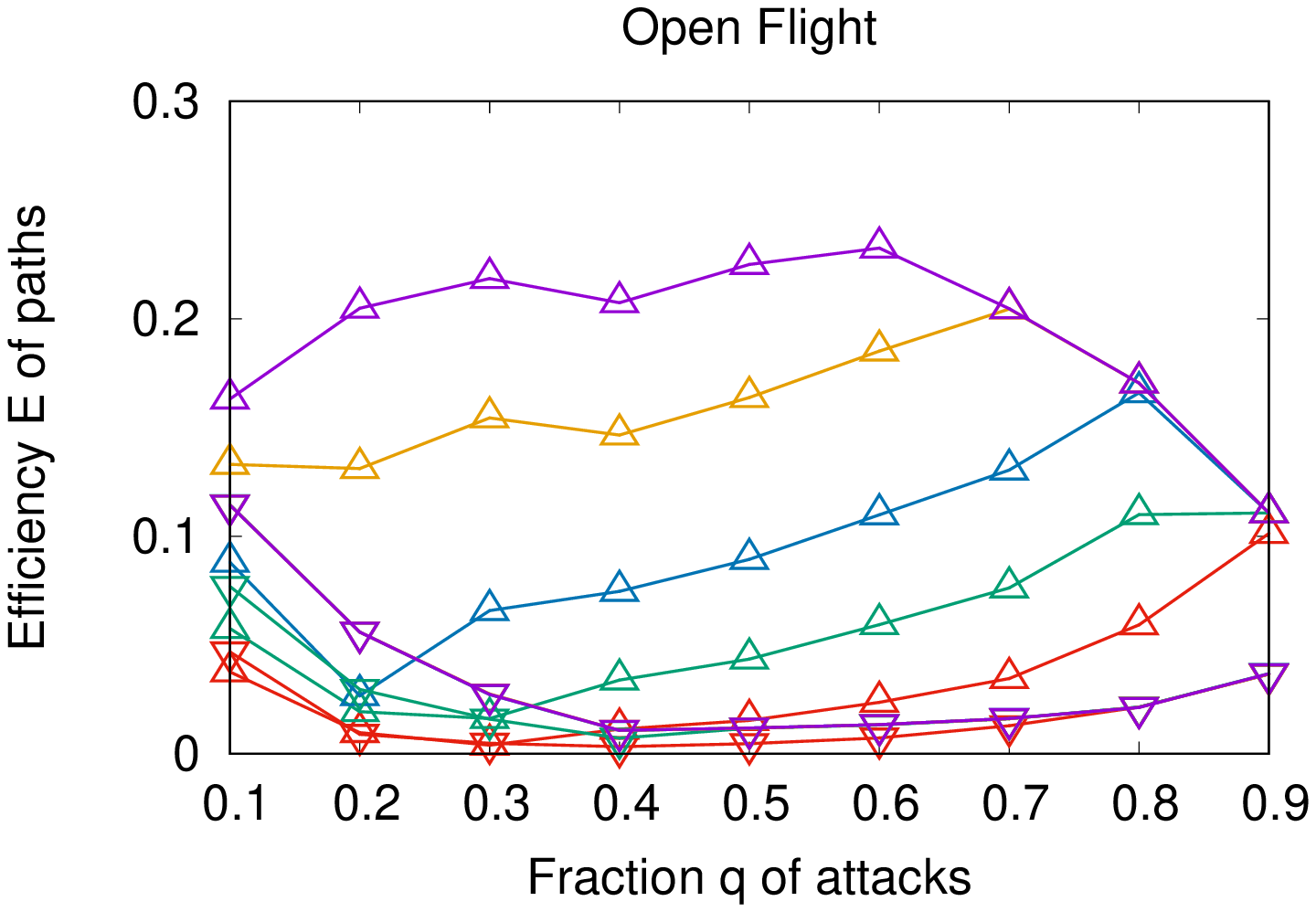}
\end{minipage}
\begin{minipage}{.49\textwidth}
\centering
  \includegraphics[width=.67\textwidth,angle=-90]{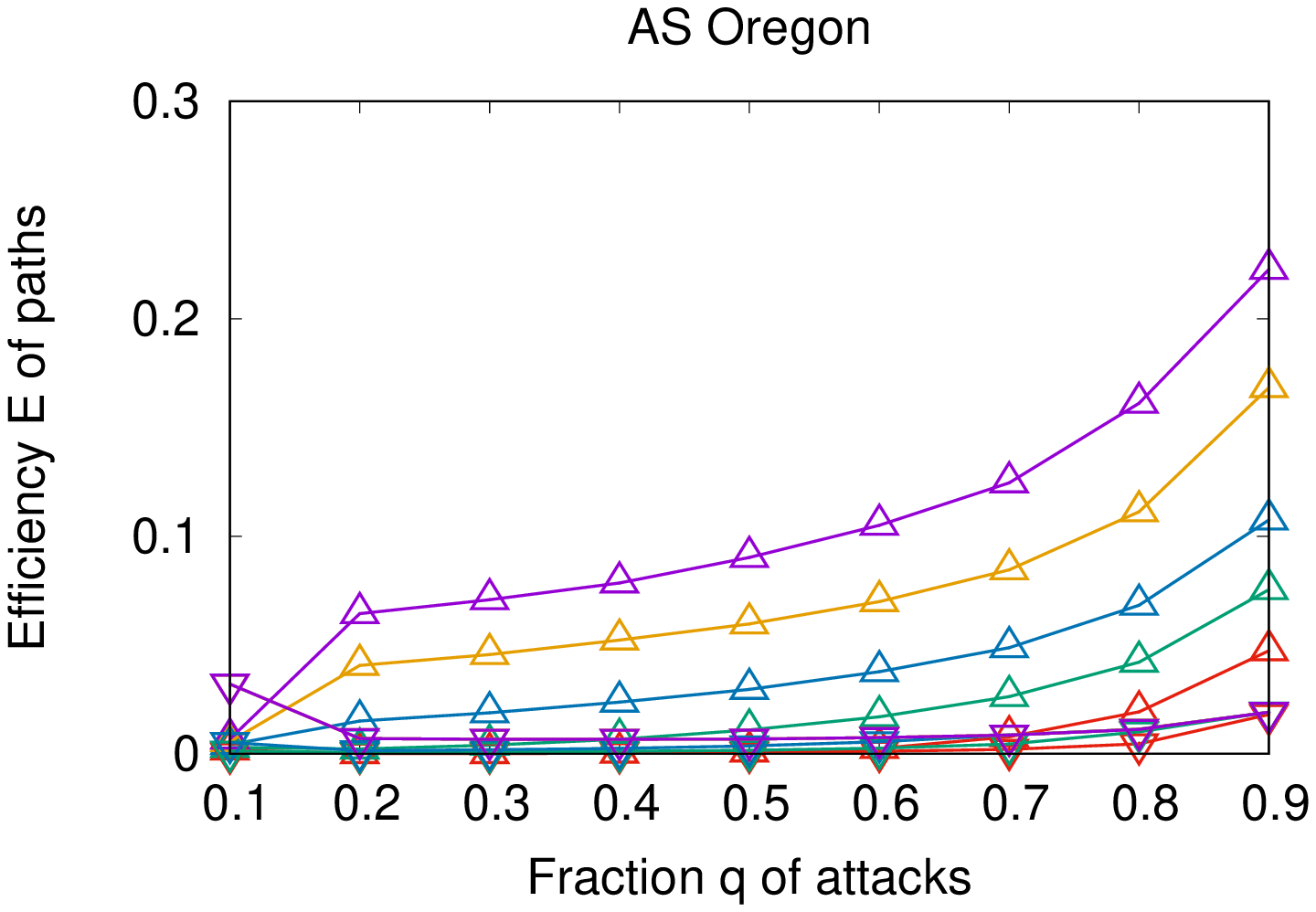}
\end{minipage}
\caption{Efficiency in the network with healing by 
only enhancing loops (up triangle) 
or extended ring (down triangle).}
\label{fig_only_E}
\end{figure}

As shown in Figs.  \ref{fig_compare_E} and \ref{fig_only_E}, 
our proposed combination method (marked by square) has 
higher efficiency than the conventional 
simple local repair method (marked by circle) in comparison 
with same color lines, 
although the effect in the method by only enhancing loops 
(marked by up-pointing triangle) or 
extended ring (marked by down-pointing triangle) 
becomes weaker with $E < 0.3$.
Dotted line shows the efficiency in the original network 
before attacks.

On the other hand, 
we consider the additional ports which should be prepared 
besides reusable ports. 
The original ports at neighbors of a removed node 
remain and can be reused, 
even if the links at the neighbor' sides are disconnected.
Thus, there exist active ports of a node 
at least as many as its degree in the original network before 
attacks.
Note that 
the minimum, average, and maximum degrees are 
$k_{min} = 1$, $\langle k \rangle = 10.77$, and $k_{max} = 242$ 
in Open Flight, 
$k_{min} = 1$, $\langle k \rangle = 3.88$, and $k_{max} = 1458$ 
in AS Oregon.
Table \ref{table_ports_proposed} shows the maximum number of 
reserved additional ports in our proposed combination method.
The number tends to be larger ranging from a few to nearly 
$2 k_{max} \sim 3 k_{max}$, 
as the fraction $q$ of attacks and the reusable rate $r_{h}$ increase.
Shown in parentheses are averaged values over the nodes that 
require additional ports in each network with healing.
The averaged number of additional ports is small, thus 
the cost of resource is so much inexpensive.
While the number is small even for the maximum 
in the conventional simple local repair method 
as shown in Table \ref{table_ports_pre92}.
It is almost constant for varying $r_{h}$ and $q$.



\section{Conclusion}
We have proposed
self-healing methods for reconstructing a sustainable network 
by rewirings against attacks or disasters in the meaning of resilience 
with adaptive capacity.
The fundamental rewiring mechanisms are based on 
maintaining the connectivity on rings and 
enhancing loops for improving the robustness by applying BP algorithm 
inspired from statistical physics.
As the resource allocation, the rewirings are controlled by a parameter 
$r_{h}$ for reuse or addition of links between the extended neighbors 
of attacked nodes.
We have shown that our proposed method is better than the 
conventional simple local repair method \cite{Gallos15} with a 
priority of rewirings to more damaged nodes, 
although reserved additional ports are required much more.
In particular, 
the whole connectivity can be revived with high efficiency of paths 
in our proposed method, when more than half links emanated from 
attacked nodes alive.
Thus, such amount of links are necessary for sustaining network 
function.
If there is lack of the resource,
the shortage parts should be compensated according to the damages.

\section*{Acknowledgment}
This research is supported in part by 
JSPS KAKENHI Grant Number JP.17H01729.

\bibliographystyle{IEEEtran}
\bibliography{bib_adaptive}

\end{document}